\documentclass[aps, notitlepage, floatfix, nofootinbib, amsmath, letterpaper,amssymb, showpacs,superscriptaddress,nobalancelastpage]{revtex4-1} 
\usepackage[colorlinks=true,
            linkcolor=blue,
            citecolor=blue,
            urlcolor=blue]{hyperref}

\usepackage{graphicx}
\usepackage{amssymb}
\usepackage{epstopdf}

\usepackage[makeroom]{cancel}

\usepackage{url}

\usepackage{geometry}

\usepackage{amstext}    % defines the \text command, needed here
\usepackage{array}      % for advanced column specification: use >{\command} for
                        \usepackage{amsmath,tabu}

\usepackage{graphicx}

\usepackage{epstopdf}

\usepackage{amsmath}

\usepackage{color}
\usepackage{xcolor}

\usepackage[makeroom]{cancel}

\usepackage{mathpazo}

\usepackage{slashed}

\usepackage{tgpagella}

\usepackage{linearA}

\usepackage{amsmath}

\usepackage{slashed}
\usepackage{mathtools}
\usepackage{ulem}

\newcommand{\ignore}[1]{}
\newcommand{\be}{\begin{equation}} \newcommand{\ee}{\end{equation}}
\newcommand{\ab}{\allowbreak}
\def\ba#1\ea{\begin{align}#1\end{align}}
\newcommand{\bit}{\begin{itemize}}
\newcommand{\eit}{\end{itemize}}
\newcommand{\im}{\item}

\newcommand{\cN}{\cal N}
\newcommand{\nn}{\nonumber} \renewcommand{\bf}{\textbf}
\newcommand{\ra}{\rightarrow} 
\renewcommand{\d}{\mathrm{d}} \newcommand{\diag}{\mathrm{diag}}
\renewcommand{\dim}{\mathrm{dim}} \newcommand{\D}{\mathrm{D}}
\newcommand{\integer}{\mathrm{integer}}

\newcommand{\tr}{\mathrm{tr}} \newcommand{\cA}{\cal A}
 
\newcommand{\cD}{\mathrm{\cal D}} \newcommand{\cF}{\cal F}
 \newcommand{\cL}{\cal L}
\newcommand{\cO}{\cal O} 
 \newcommand{\s}{\,\,\,}
\renewcommand{\a}{\alpha} \renewcommand{\b}{\beta}
\newcommand{\e}{\mathrm{e}} \newcommand{\eps}{\a}
\newcommand{\f}{\phi} \newcommand{\fr}{\frac} \newcommand{\g}{\gamma}
\newcommand{\h}{\hat} \renewcommand{\i}{\mathrm{i}}
\newcommand{\p}{\partial}  \newcommand{\x}{\xi}
\newcommand{\EE}{\vec E}  \newcommand{\NN}{\vec \nabla}

\def\slashb#1{\setbox0=\hbox{$#1$}#1\hskip-\wd0\dimen0=5pt\advance
        \dimen0 by-\ht0\advance\dimen0 by\dp0\lower0.5\dimen0\hbox
          to\wd0{\hss\sl/\/\hss}}
\def\bra#1{\left< #1\right|}
\def\ket#1{\left| #1\right>}
\def\bracket#1#2{\left<#1\mid #2\right>}
\def\EV#1#2#3{\bra{#1}#2\ket{#3}}
\input{epsf}
 \newcommand{\eight}        {$\sqrt{s}~=~8$~Te\kern-.1emV\xspace}
 \newcommand{\thirteen}        {$\sqrt{s}~=~13$~Te\kern-.1emV\xspace}

\begin{document}
\def\xy{(x, \, y)}
\def\xyz{(x, \, y, \, z)}
\def\cH{\cal H}
\def\H{Hamiltonian }
\def\EL{\cal{EL}}
\def\eq{\end{quotation}}
\def\bq{\begin{quotation}}

\def\bqs{{\small \begin{quotation}}
\def\eqs{\end{quotation}}}
\raggedbottom

\def\pex{\paragraph{Exercise:}}

\def\bq{\quotation}

\def\ab{\allowbreak}
\def\nn{\nonumber}
\def\ra{\rightarrow}
\def\d{\mathrm{d}}
\def\bf{\textbf}
\def\diag{\mathrm{diag}}
\def\dim{\mathrm{dim}}
\def\D{\mathrm{D}}
\def\E{\mathrm{E}}
\def\integer{\mathrm{integer}}
\def\tr{\mathrm{tr}}
\def\cA{\cal A}
\def\cD{\mathrm{\cal D}}
\def\cF{\cal F}
\def\cL{\cal L}
\def\cO{\cal O}
\def\s{\,\,\,}
\def\a{\alpha}
\def\b{\beta}
\def\e{\epsilon}
\def\f{\phi}
\def\fr{\frac}
\def\g{\gamma}
\def\h{\hat}
\def\i{\index}
\def\p{\partial}
\def\x{\xi}
\def\AA{\vec A}
\def\AAt{\underline{\vec A}}
\def\AAl{ \vec A_{{\cal L}}}

\def\BB{\vec B}
\def\EE{\vec E}
\def\NN{\vec \nabla}
\def\JJ{\vec j}
\def\DD{\Delta}
\def\xx{\vec x}
\def\eps{\epsilon_{0}}

\def\ba#1\ea{\begin{align}#1\end{align}}

\def\bit{\begin{itemize}}
\def\eit{\end{itemize}}
\def\im{\item}

\newcommand{\mm}[1]{\marginnote{#1}}
\newcommand{\nnn}{\nn \\ &}

\newcommand{\sch}{Schr\"{o}dinger }

\def\N{Newtonian }
\def\cN{{\cal N}}
% four Dirac
\def\bra#1{\left< #1\right|}
\def\ket#1{\left| #1\right>}
\def\bracket#1#2{\left<#1\mid #2\right>}
\def\EV#1#2#3{\bra{#1}#2\ket{#3}}

\title{Mirror Quantum Tomography Finds Unexpected Polarization Phenomena in $Z$ Boson Production in pp Collisions at the LHC}
\author{A. Gautam}
\affiliation{Department of Physics and Astronomy, The University of Kansas, Lawrence, KS 66045, USA\vspace{0ex}}
\author{J. C. Martens}
\affiliation{Department of Physics and Astronomy, The University of Kansas, Lawrence, KS 66045, USA\vspace{0ex}}
\author{J. P. Ralston}
\email{ralston@ku.edu}
\affiliation{Department of Physics and Astronomy, The University of Kansas, Lawrence, KS 66045, USA\vspace{0ex}}
\author{G. Stejskal}
\affiliation{Department of Physics and Astronomy, The University of Kansas, Lawrence, KS 66045, USA\vspace{0ex}}
\author{J. D. Tapia Takaki}
\email{daniel.tapia.takaki@cern.ch}
\affiliation{Department of Physics and Astronomy, The University of Kansas, Lawrence, KS 66045, USA\vspace{0ex}}
\begin{abstract}
{\it Abstract: The ATLAS, CMS and LHCb collaborations have reported data for the inclusive production of lepton pairs with invariant mass 80-100 GeV in pp collisions. We find the ATLAS data from  at \eight shows an unexpected vortex-like configuration of $Z$ boson spins circulating around the beam axis. Associated with this structure is a local maximum of the entropy of the $Z$-boson polarization density matrix extracted using quantum tomography. Data from CMS and LHCb at \thirteen are broadly consistent. The origin of the observed phenomena is unknown but generally related to observables that are formally odd under charge conjugation, parity and time-reversal. Quantum tomography using Standard Model lepton couplings to the $Z$ boson determines the spin-1 density matrix of the system model-independently, bypassing any model of the hadronic production process.}
\end{abstract}

%Recently the ATLAS Collaboration reported data for lepton pairs in pp collisions at \eight in the region of invariant mass 80-100 GeV. They note broad agreement with Standard Model predictions, along with puzzling discrepancies that were associated to QCD. Results from the CMS Collaboration and the LHCb Collaboration in pp collisions at \thirteen are also available and are broadly consistent with those reported by the ATLAS Collaboration. By re-analyzing the data utilizing quantum tomography, we observe a previously unreported spin-vortex structure in the Z boson production, indicating the presence of an additional mechanism. }

\maketitle

It is generally believed that production of inclusive $Z$ bosons in $pp$ collisions can be computed with high accuracy in perturbative QCD. The distribution of inclusive lepton pairs with invariant mass ${80-100} \, GeV$ in such collisions is believed to be dominated by $Z$ boson production. The ATLAS~\cite{ATLAS:2016rnf} collaboration measured the angular distribution of dileptons in that region and rapidity ${| y | < 3.5}$ from $pp$ collisions with center-of-mass energy \eight. We have discovered the ATLAS data shows a vortex-like configuration of vector boson spins, plus other unexpected polarization observables, which may be evidence for non-trivial dynamical features of the production process. Measurements of dilepton angular distributions have also been reported by the CMS Collaboration and the LHCb Collaboration~\cite{CMS:2015cyj,LHCb:2022tbc}. This work uses the measurements of the ATLAS Collaboration~\cite{ATLAS:2016rnf} as a reference, while including CMS and LHCb results to illustrate the overall agreement among three experiments.

Our study began as an application of {\it quantum tomography} to collider physics, first introduced in Ref.~\cite{Martens:2017cvj}. We initially assumed that $Z$ bosons would represent a basic test case for a new method. Tomography uses a known probe density matrix $\rho_{probe}$ to determine an unknown density matrix $\rho_{X}$. When quantum tomography was introduced by Fano~\cite{Fano:1957zz} it was not appreciated. Its power was later appreciated in quantum optics~\cite{Smithey:1993zz} and many other fields to become a mainstream method to reconstruct a quantum state~\cite{Wootters:1989lkz,Gross:2010aat,Lundeen:2011mnu,Kueng:2014bkr, Haah:2015rrj,ODonnell:2016ume}. 

The traditional application of quantum field theory to scattering processes sets up an over-complete framework with many unobservable elements. Inclusive processes will typically parameterize hadronic matrix elements as unknown "structure functions", which may or may not be experimentally accessible. The approach here~\cite{Martens:2017cvj} concentrates directly on {\it the observables\footnote{We never use the word ``observable" to mean an operator. An observable is a number an experiment can observe.} of a given experiment}, bypassing unobservable formalism. An unknown quantum mechanical system is characterized by its density matrix $\rho_{X}$. From quantum mechanics, the expectation value of a set of operators called $\rho_{probe}$ is $tr(\rho_{X}\rho_{probe})$. The expression is the Hilbert-Schmidt inner product on the space of operators. 

The particular terms observable in $\rho_{X}$ are strictly limited to the independent terms of the probe. The formalism only refers to what is observable. Adopting this as a {\it matter of principle} transforms the interpretation of quantum tomography, $QT$. We deny a standard interpretation that $QT$ is exponentially difficult, which is based on the exponentially high dimensionality of composite quantum systems. We accept a new  interpretation that a density matrix is {\it defined to represent what can be observed}. The key conceptual viewpoint of mirror tomography was originally introduced in Ref.~\cite{Martens:2017cvj}, which writes ``the observable target structure is always a mirror of the probe structure." The mirror property is exact, inasmuch as the Hilbert-Schmidt inner product localizes particular information in mutually orthogonal subspaces which are 1-1 linearly connected between experimental detection and theoretical description. The mirror property has begun to appear under a variety of names~\cite{Aaronson:2017qlb,Huang:2020tih} in the literature. More review  of quantum tomography in high energy physics can be found in~\cite{Martens:2017cvj,Martens:2019bxv,Martens:2022wtb,Ralston:2022snp}, along with subsequent related studies~\cite{Afik:2020onf,Afik:2022kwm,Aoude:2022imd,Ashby-Pickering:2022umy,Altomonte:2023mug,Fabbrichesi:2023jep,Dong:2023xiw,Bernal:2023jba,Aguilar-Saavedra:2024vpd,Fabbrichesi:2024wcd,Cheng:2024rxi,Ruzi:2024iqu,Altomonte:2024upf,Aguilar-Saavedra:2025byk,DelGratta:2025xjp,Ruzi:2025jql}. The method avoids making any model of the unknown system $\rho_{X}$, while using information known about the probe. 

The probe of this analysis is the dilepton polarization density matrix of the Standard Model with leading order vector and axial coupling vertices to the $Z$-boson. While we will call the system "a $Z$ boson" for reference, $Z-\gamma$ interference and any other production process {\it of the system being probed} is automatically taken into account. Also taken into account are all radiative corrections to the production and final state lepton observables, as well as all detector effects not removed from the data set. No assumptions are made about the preparation of the hadronic state, nor about parton distributions, nor perturbative sub-processes such as initial state radiation. The analysis takes into account what can be observed without bias of interpretation. Our illustration with the relatively simple probe of a vector boson system represents a very tiny fraction of experiments the method is capable of exploring with more general probes.

In the experiment, two hadrons with 4-momenta $P_{A}$, $P_{B}$ inclusively produce lepton-antilepton pairs with 4-momenta $k$ (positive charge), $k'$ (negative charge), with $k^{2} \sim k^{'2} \sim 0$ in the high energy limit. Let the total pair momentum  $Q=k+k'$, which has components $Q^{\mu} = (\sqrt{Q^{2}}, \, \vec Q=0)$ in its own rest frame. The difference momenta $\ell =k-k'$ satisfy $\ell_\mu Q^\mu=0$. The components of $\ell_\mu =\sqrt{Q^{2}} \hat \ell$ are entirely spatial (3-vectors) in the rest frame of $Q.$ The unit vector $\hat \ell$ is characterized by polar and azimuthal angles $\theta, \, \phi$ in the same frame. Then the dilepton phase space is \ba k_{0}k_{0}' {d\sigma \over d^{3}k  d^{3}k'} = {d \sigma \over  d^{4}Q d\Omega } ={dN(\theta, \,\phi \big | Q) \over d\cos \theta d \phi}{d\sigma \over d^{4}Q }\nn \ea Here $dN(\theta, \,\phi \big| Q)/d\Omega$ is the angular distribution conditional upon $Q$, which will be our topic. 

Dilepton angles are defined event by event in terms of a set of three {\emph frame 4-vectors} $X^{\mu}, \, Y^{\mu}, \, Z^{\mu}$. The set defines a spacelike coordinate frame akin to $xyz$ in the $\vec Q=0$ frame. Mutual orthogonality requires \ba & Q\cdot X=Q\cdot Y =Q\cdot Z=0; \nn \\ & X\cdot Y=Y \cdot Z =X\cdot Z=0.  \label{Qdot} \ea Defining $P_{A}$=(1, 0, 0, 1), $P_{B}$=(1, 0, 0, -1) (light-cone $\pm$ vectors), the Collins-Soper frame satisfying the relations of Eq. \ref{Qdot} is given by \ba \tilde Z^{\mu} &=  P_{A}^{\mu} Q\cdot P_{B}  -  P_{B} ^{\mu} Q\cdot  P_{A}; \nn \\ \tilde X^{\mu} &= Q^{\mu}- P_{A}^{\mu} {Q^{2}\over 2 Q\cdot P_{A}} - P_{B} ^{\mu} {Q^{2}\over 2 Q\cdot P_{B} }; \nn \\ \tilde  Y^{\mu}&= \epsilon^{\mu \nu \a \b}P_{ A\nu}P_{B \a}Q_{\b}. \nn \ea The normalized basis is \ba (X^{\mu}, \, Y^{\mu}, \, Z^{\mu}) =  ({\tilde X^{\mu} \over \sqrt{-\tilde X\cdot \tilde X} }, \, {\tilde Y^{\mu} \over \sqrt{-\tilde Y\cdot \tilde Y} }, \, {\tilde Z^{\mu} \over \sqrt{-\tilde Z\cdot \tilde Z} }) . \label{basis} \ea Here $\epsilon^{0123}=1$. This is the Collins-Soper ($CS$) frame for light-like beam particles.

Expressing frame 4-vectors Lorentz covariantly is very efficient. For each event $J$, the expansion of the dilepton angular variables $\hat \ell_{J}$ in the basis of Eq. \ref{basis} is \ba \hat \ell_J =\left(\hat \ell_J^\mu X_\mu, \, \hat \ell_J^\mu Y_\mu,\, \hat \ell_J^\mu Z_\mu, \right) = (\sin \theta \cos \phi, \,\sin \theta \sin \phi, \,\cos \theta  )_J . \ea No Lorentz boosts are involved in the calculation, which is conveniently done with lab-frame 4-momenta. This method is presented in Ref.~\cite{Martens:2017cvj}, and it has been discussed in several subsequent works~\cite{Gavrilova:2019jea}. 
 
Event by event, the density matrix of lepton unit vectors coupled to a Z-boson is \ba \rho_{ij}(\ell) = {1\over 3}\delta_{ij}  +\sin^{2}\theta_{W} \hat \ell \cdot \vec J_{ij} +{1\over 2} U_{ij}(\hat \ell) , \qquad \text{where} \quad U_{ij}(\hat \ell)= {\delta_{ij} \over 3} -\hat \ell_{i}\hat \ell_{j}.  \label{lastline} \ea Here $(\vec J_{ij})_{k}= -\imath \epsilon_{ijk}$ are the spin-1 rotation generators in Cartesian form. Each term proportional to 1, $\vec J$ and $U(\hat \ell)$ is orthogonal to the others, while $U(\hat \ell)$ can be further decomposed into 5 orthogonal matrices. 

By the mirror property the most general unknown system density matrix (denoted $\rho(Z)$) that can be observed with the lepton probe has the same general expansion: \ba \rho_{ij}(Z) = {1\over 3}\delta_{ij}  +{1\over 2}\vec S(Z)\cdot \vec J_{ij} +U_{ij}(Z),\qquad \text{where} \quad U(Z)=U^{T}(Z); \quad tr(U(Z))=0.\label{nextline} \ea Here $\vec S(Z)$ stands for the {\it spin parameters} of the $Z$-boson density matrix. A pure state with $|\vec S|=1$ has a density matrix \ba \rho_{pure, \, ij}(\vec S) ={1\over 3}1+{ 1 \over 2}\vec S\cdot \vec J. + {1\over 2} ({1\over 3}\delta_{ij}- \hat S_{i}\hat S_{j})  \label{pure} \ea Unlike the case of spin-1/2, pure states with $\vec S=0$ exist. The traceless symmetric matrix of parameters $U_{ij}(Z)$ is called the quadrupole (spin-2) term and contributes to the mixed-state {\it degree of polarization} and {\it entropy} that will be discussed shortly.

It is convenient to expand $U_{ij}(\hat \ell)$ in a basis of traceless orthonormal matrices dual to real spin-2 spherical harmonics $Y_{M}(\theta, \, \phi)$. The spin-2 contributions to the angular distribution are then $tr(\rho(\ell) \rho(Z))_{spin-2} \sim \sum_{M} \, \rho_{M}(Z) Y_{M}(\theta, \,  \phi)$. Calculation gives \ba  {dN \over d \Omega}=& {1\over 4 \pi}+{\frac{3}{4 \pi}}S_{x}\sin\theta \cos\phi+ {\frac{3}{4 \pi}}S_{y}\sin\theta \sin\phi+{\frac{3}{4 \pi}}S_{z}\cos\theta  \nn \\ &+c\rho_{0}  {({1\over \sqrt{3}} -  \sqrt{3} \cos^2\theta)}  {-} c \rho_{1} \sin(2 \theta)  \cos\phi 
+ c\rho_{2}  \sin^{2}\theta\cos(2 \phi) \nn \\ & +c\rho_{3} \sin^{2}\theta  \sin(2 \phi) {-} c \rho_{4} \sin(2 \theta)  \sin \phi .\label{res1} \ea Here $c= 3 /(8\sqrt{2} \pi ) $ and the label $Z$ has been suppressed.~\footnote{An open-source code of the mirror quantum tomography is available at \url{http://cern.ch/quantum-tomography}}

\begin{small} 

\begin{table*}[ht]
\centering

$\begin{array}{|lll || c | c | c | c | c |}
\hline
term & origin & dN/d\Omega & C_{\ell}  & P & T & C_{\ell}P &PT \\ 
\hline 
 \cdot & \ell & \cdot & -  &   - & -   & +  & +  \\  
 \cdot &  X & \cdot  & \cdot  &   -& -   &+   & + \\   
 \cdot  &  Y &\cdot  &  \cdot  &  + & +  &+    & +  \\ 
 \cdot  &   Z &  \cdot  &  \cdot    & -  & -   & +   & + \\ 
S_{x}&   X\ell &  \sin\theta \cos \phi  &-  & +  & +  & -  & +\\ 
S_{y}  &  Y\ell  &  \sin\theta \sin \phi  & - & -  & -   & + & + \\ 
S_{z}  &   Z\ell  &  \cos\theta    &  - & +  &+   & -  & + \\ 
\rho_{2} &   XX \ell \ell & \sin^{2}\theta\cos 2\phi    &    + &  + & +   &  +  &+ \\ 
\rho_{3}   &    XY \ell \ell   &  \sin^{2}\theta\sin2\phi      &  +  &  - &  - &  +  & + \\ 
\rho_{1}   &     XZ \ell \ell  & \sin 2\theta \cos\phi        &    + &  + & +   &  +  & + \\ 
\rho_{4}     &   YZ \ell \ell  &  sin 2\theta\sin\phi      &  +  &  - &  - &  +  & + \\ 
\rho_{0}     &    ZZ\ell \ell   & 1/\sqrt{3} -\sqrt{3} \cos^{2}\theta     &    + &  + & +   &  +  & + \\  \hline 
  \end{array}$

  \caption{ \small Terms in the angular distribution classified under discrete transformations $C_{\ell}$, $P$, and $T$. Here $\ell$ stands for $\hat \ell$, $ X\ell $ stands for $\hat X \cdot \hat \ell=-X_{\mu} \ell^{\mu}$, and so on with scalar normalization factors removed. $T$-odd scattering observables from imaginary parts of amplitudes generally exist without violating fundamental $T$ symmetry. See the text for more explanation. }\label{tab:symmetries}
\end{table*}   \end{small} 

Many studies including ATLAS report angular distribution coefficients in a convention $A_{0} -A_{7}$ assigned without physical interpretation. By inspection $(A_{3} , \, A_{7}, \, A_{4})/4 =(S_{x} , \, S_{y}, \, S_{z})$. Table \ref{tab:symmetries} lists discrete transformation properties of all terms under parity $P$, time reversal $T$, and
lepton charge conjugation $C_{\ell}$. Notice that $S_{y}$ is odd under $C_{\ell}$, $P$, and $T$. In scattering $T-$odd quantities such as $S_{y} \neq 0$ are not directly related to violation of fundamental time reversal symmetry. $T-$odd behavior is theoretically noted~\cite{GAMBERG2007362} for being connected to imaginary parts of scattering amplitudes. Both loop corrections in perturbation theory and physics beyond the Standard Model can produce $T$-odd terms~\cite{Alioli:2020kez,Petriello:2025lur}. The complex amplitude structure of $Z$ production is first probed in perturbative $QCD$ at next-to-next to leading order ($NNLO$).

\begin{figure}[ht]
\begin{center}
\includegraphics[width=4in]{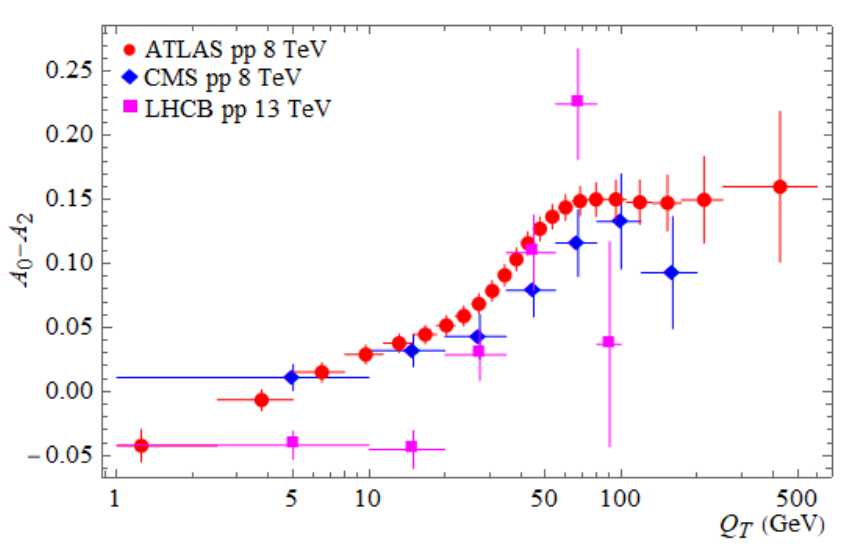}
\caption{The difference $A_{0}-A_{2}$ of the angular coefficients as a function of the Z-boson transverse momenta, and measured by the ATLAS, CMS and LHCb collaborations in proton-proton collisions. The error bars show the total uncertainty reported by the experiments.}
\label{fig:lastline}
\end{center}
\end{figure}

\begin{figure}[ht]
\begin{center}
\includegraphics[width=4in]{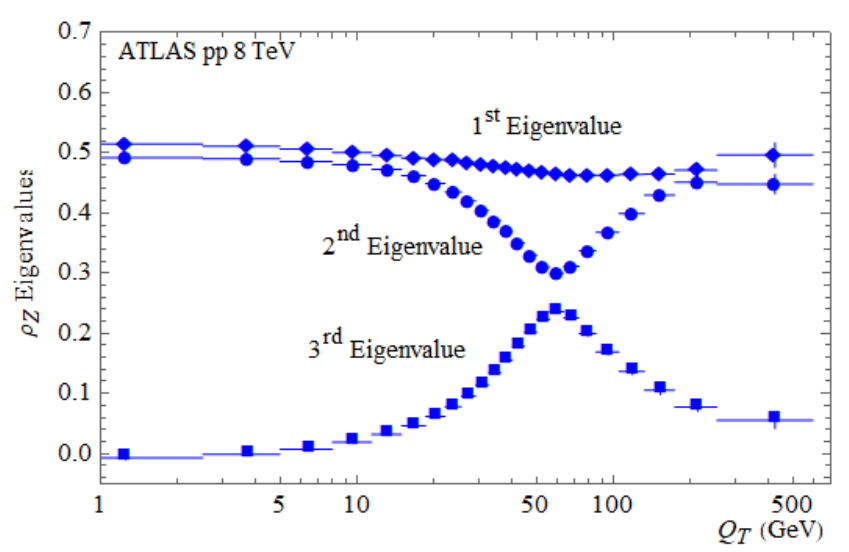}
\includegraphics[width=4in]{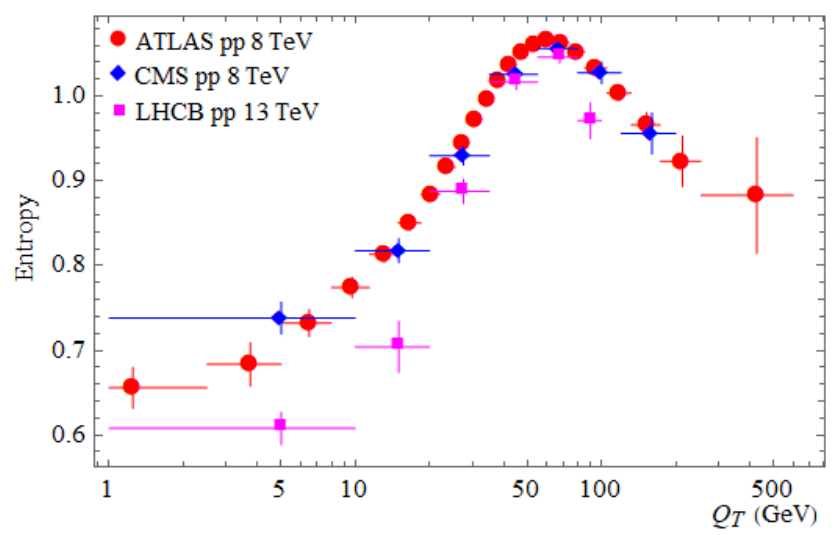}
\includegraphics[width=4in]{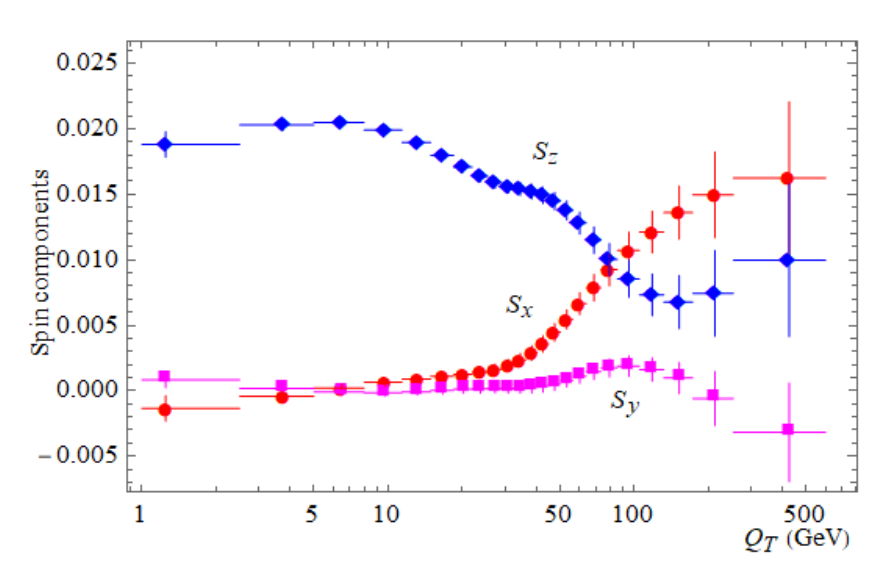}
\caption{Polarization density matrix eigenvalues, von Neumann entropy, and spin components obtained from LHC data. For simplicity ATLAS eigenvalues and the spin components are shown, with the entropy of the ATLAS, CMS and LHCb combined. Statistical and systematic uncertainties of the experimental measurements have been added in quadrature.}
\label{fig:fig-results}
\end{center}
\end{figure}

The $\vec Y$ direction is special. Event by event, the $\vec Z$ direction in the lab frame is essentially the (rapidity signed) beam momentum, which is odd under $T$. The $\vec X$ direction in the lab frame is essentially the pair transverse momentum $\vec Q_{T}$ up to corrections for orthogonality. It is also odd under $T$. The $\vec Y$ direction is along $\vec Z \times \vec X$, making it a pseudovector under parity and {\it even} under $T$, so that $S_{Y}= \vec S\cdot \hat Y$ {\it T-odd.} The traditional labels $X$ and $Y$ are a misnomer. The $\hat X$ direction is more like a radial cylindrical coordinate, and the $\hat Y$ direction point-by-point corresponds to the direction of increasing azimuthal angle $\hat \phi$. The relation is $\hat \rho = cos\phi \hat X+ sin\phi \hat Y;  \, \hat \phi =-sin\phi \hat X+ cos \phi \hat Y$. The corresponding Cartesian frame fixed in the lab is then related by a Lorentz boost. The transverse components of vectors are unchanged by a CM-lab boost parallel to the lab frame $z$ axis. 

As an example of the cross-experiment comparison, Figure~\ref{fig:lastline} shows the difference $A_{0}-A_{2}$ of the angular coefficients as a function of the Z-boson transverse momentum. The data of CMS and LHCB are compatible with ATLAS, which consistently reports smaller uncertainties. 

Figure \ref{fig:fig-results} shows an unexpected resonant-type behavior in the $Z$-boson density matrix eigenvalues as a function of the pair transverse momentum $Q_{T}$. Near $Q_{T} \sim 65$ GeV the eigenvalues nearly become degenerate. Also shown is $|\vec S|$, which undergoes a significant dip in the same vicinity. We do not have an explanation of these phenomena, nor the $Q_{T}$ scale at which they occur.

The middle panel of Figure \ref{fig:fig-results} shows the entropy $S = -tr(\rho_{Z} \log(\rho_{Z}))$ across the same region. It has a local maximum, which is characteristic of an increased dynamical phase space.\footnote{Positivity of a density matrix requires eigenvalues in the range zero to one. Data fitting procedures not implementing positivity can violate the requirement, which produces the low-$Q_{T}$ cutoff visible in the entropy. Ref~\cite{Martens:2017cvj} discusses practical methods to maintain positivity in data analysis.} The maximum observed $S \sim 1.06$ can be compared to the maximum possible entropy $\log(3) =1.0986$ for a system with three eigenvalues. When entropy is maximal the entropy of polarization is zero: Such a boson is completely {\it unpolarized.}

\begin{figure}[ht]
\begin{center}
\includegraphics[width=4in]{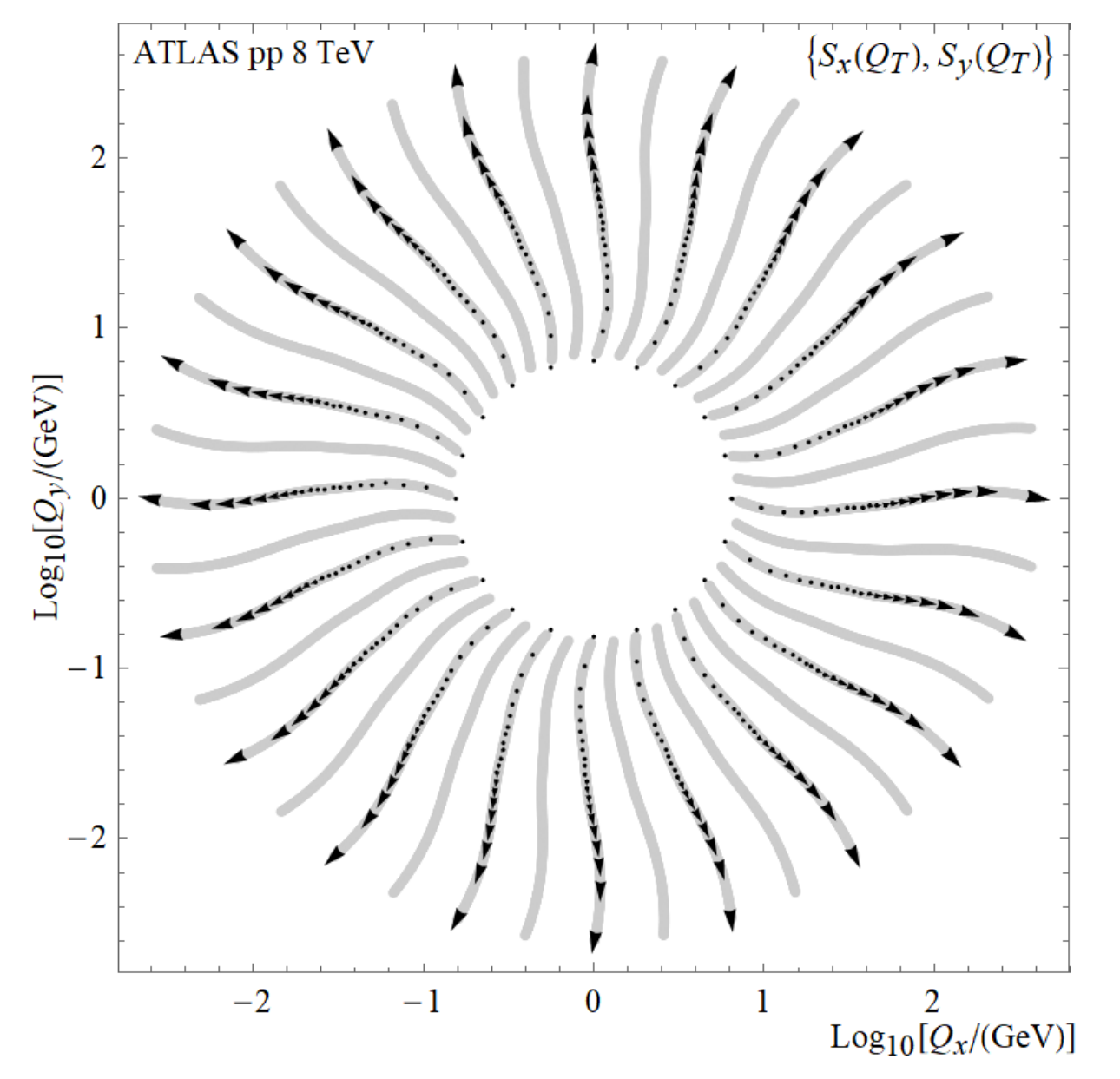}
\caption{The spin vector field $(S_{x}(Q_{T)}, \, S_{y}(Q_{T})) $. The view looks down the $Z$ axis at the transverse plane with radial coordinate $log_{10}(Q_{T}/GeV).$ The tail of arrows are placed at each $Q_{T}$ value. The curves indicate streamlines tangent to the vector field. }
\label{fig:SpiralNoScale}
\end{center}
\end{figure}

% \end{figure}

%\label{fig:SpiralBowl}
%\end{center}
%\end{figure}

The bottom panel of Figure \ref{fig:fig-results} shows the components of $\vec S$ as a function of $Q_{T}$ multiplied by 10. This plot also shows dramatic resonance-like variation in the region of $Q_{T} \sim 90-110$ GeV. Notice that $S_{Y}$ goes through a maximum in the same region and changes sign at large $Q_{T}$. 

Production in the lab frame is azimuthally symmetric with respect to the beam axis. The event-by-event distribution is repeated at arbitrary laboratory azimuthal angles, creating the structure seen in Figure~\ref{fig:SpiralNoScale} showing the transverse components of $\vec S$. The perspective looks down the $Z$ axis at the transverse plane with radial coordinate $log_{10}(Q_{T}/GeV)$ to separate the arrows visually. The arrows show the orientation of $\vec S_{T}$ in fixed coordinates $\hat x, \, \hat y$ relative to $\vec Q_{T}$ and along a fixed azimuthal angle $\phi$. That is, \ba \vec S = S_{X} \hat \rho +S_{Y} \hat \phi = \left( S_{X} cos\phi -S_{Y} sin \phi \right) \hat x + \left( S_{X} sin\phi + S_{Y}  cos \phi \right) \hat y. \nn  \ea The rest of the figure is made from copies related by azimuthal symmetry. The ensemble configuration of $Z$-spins thus produces a spiral vector field $S(\vec Q)$ in the lab frame. The sense of rotation of the vortex is direct visual evidence for both parity and charge conjugation symmetry violation. 

It is interesting to re-examine Figure \ref{fig:fig-results} which shows the $X$ (effectively radial) and $Z$ (longitudinal) components of $\vec S$ tend to be much larger than $S_{y}$. In three dimensions the $Z$-spins are aligned along a twisted bowl-shaped structure. The symmetry of $pp$ collisions requires the bowl be reversed on opposite sides of the beam, with the same odd$-P$ sense of chirality.

In summary, the ATLAS data for dilepton production with the invariant mass of $Z$ bosons shows unexpected dynamical behavior reminiscent of resonances and coherent quantum mechanical phenomena. Any temptation to interpret this as a signal of new physics must be tempered by the overall trend of the data to agree with Standard Model predictions. Yet ATLAS has noted significant discrepancies with pQCD predictions at $NNLO$ order~\cite{Anastasiou:2003yy}, including the generalized Lam-Tung relation. That relation is intimately connected to the reality of amplitudes at tree level. Violation of the relation is connected to imaginary amplitudes that are a challenge for pQCD. The unexpected spin-vortex structure is directly connected to the puzzle by the $T$-odd character of $S_{y}$. A more definite interpretation of these new phenomena appears to be premature at this time. It would be interesting to look for spin vector field structure in other processes, such as the inclusive production of charmonium, bottomonium, dijets, $HH$, $W^+ W^-$, $ZZ$, and ``boosted" top quark production. 

\begin{acknowledgments}
This work was partly supported by the U.S. Department of Energy Office of Science through the DOE Nuclear Physics and EPSCoR programs. 
\end{acknowledgments}

\section*{References}
\bibliography{2026biblio}

\end{document}